\documentclass[preprint,aps,nofootinbib]{revtex4}
\usepackage{}
\usepackage{graphicx}
\usepackage{amsmath}
\usepackage{amsfonts}
\usepackage{mathrsfs}
\usepackage{amssymb}
\usepackage{color}%
\usepackage{dcolumn}
\definecolor{lightgray}{rgb}{.7,.7,.7}

\definecolor{red}{rgb}{1,0,0}

\definecolor{green}{rgb}{0,1,0}

\definecolor{blue}{rgb}{0,0,1}

\providecommand{\U}[1]{\protect\rule{.1in}{.1in}}

\newcommand{\f}{\begin{equation}}
\newcommand{\ff}{\end{equation}}
\newcommand{\fa}{\begin{eqnarray}}
\newcommand{\ffa}{\end{eqnarray}}

\begin{document}

\title{The Petrov-like boundary condition at finite cutoff surface in Gravity/Fluid duality}
\author{Yi Ling $^{1,2,3}$}
\email{lingy@ihep.ac.cn}
\author{Chao Niu $^{1}$}
\email{niuc@ihep.ac.cn}
\author{Yu Tian $^{4,2}$}
\email{ytian@ucas.ac.cn}
\author{Xiao-Ning Wu $^{5,6,2}$}
\email{wuxn@amss.ac.cn}
\author{Wei Zhang $^{3,1}$}
\email{weizhangncu@163.com}

\affiliation{$^1$Institute of High Energy Physics, Chinese Academy
of Sciences, Beijing 100049, China\\ $^2$ State Key Laboratory of
Theoretical Physics, Institute of Theoretical Physics, Chinese
Academy of Sciences, Beijing 100190, China\\ $^3$ Center for
Relativistic Astrophysics and High Energy Physics, Department of
Physics, Nanchang University, 330031, China\\
$^4$ School of Physics, University of Chinese Academy of Sciences,
Beijing 100049, China\\
$^5$Institute of Mathematics, Academy of Mathematics and System
Science, Chinese Academy of Sciences, Beijing 100190, China\\
$^6$Hua Loo-Keng Key Laboratory of Mathematics, CAS, Beijing
100190, China}

\begin{abstract}
Previously it has been shown that imposing a Petrov-like boundary
condition on a hypersurface may reduce the Einstein equation to
the incompressible Navier-Stokes equation, but all these
correspondences are established in the near-horizon limit. In this
paper, we demonstrate that this strategy can be extended to an
arbitrary finite cutoff surface which is spatially flat, and the
Navier-Stokes equation is obtained by employing a non-relativistic
long-wavelength limit.
\end{abstract} \maketitle

\section {Introduction}
It has been known that the excitations of a black hole horizon
dissipate very much like those of a fluid since the 70's of last
century\cite{Hawking:1972hy,Damour1979,Damour1982,Price:1986yy}.
From then on, the gravity/fluid duality has been heavily
investigated  and lots of important progress have been
made\cite{Jacobson:1995ab,Policastro:2001yc,Policastro:2002se,
Kovtun:2003wp,Buchel:2003tz,Bhattacharyya:2008jc,
Bhattacharyya:2008xc,Bhattacharyya:2008ji,Banerjee:2008th,
Iqbal:2008by,Bhattacharyya:2008mz,Bhattacharyya:2008kq,
Hansen:2008tq,Caldarelli:2008ze,Eling:2009pb,Bredberg:2010ky,
Heemskerk:2010hk,Faulkner:2010jy,
Padmanabhan:2010rp,Bredberg:2011jq,Compere:2011dx,Hu:2011ze,
Cai:2011xv,Lysov:2011xx,MT,Niu:2011gu,Huang:2011he,Huang:2011kj,
Compere:2012mt,Eling:2012ni,Cai:2012vr,Matsuo:2009pg,Zhang:2012uy,Cai:2013uye,Wu:2013kqa,
Wu:2013mda,Tian:2012bg,Zou:2013fua}. Remarkably, recent progress
in AdS/CFT correspondence has shed more insightful light on this
duality. The method of hydrodynamical expansion of the metric was
initially proposed to study the dual fluid living on the boundary
of spacetime, in which the regularity condition is imposed on the
horizon and a long-wavelength expansion is
needed\cite{Bredberg:2011jq}. Later, an alternative way was
proposed to reduce the Einstein equation to the Navier-stokes
equation by imposing a Petrov-like boundary condition on the
cutoff surface\cite{Lysov:2011xx}. The key idea of this strategy
is to consider the perturbations of the extrinsic curvature of the
cutoff surface directly, rather than those of the metric. As a
result, the Brown-York stress tensor is treated as the fundamental
variable which due to the holographic dictionary can be identified
with the stress-energy tensor of a fluid living on the cutoff
surface\cite{Bredberg:2011jq,Tian:2012bg}. In another word, we may
extract the hydrodynamical behavior of gravity directly, needless
to solve the perturbation equation for the explicit form of the
perturbed metric. In literature the advantages of this strategy
have been continuously disclosed. It has been successfully applied
to a spacetime with a spatially curved cutoff
surface\cite{Huang:2011he}, or a spacetime with a cosmological
constant as well as matter fields\cite{Huang:2011kj,Zhang:2012uy}.
In particular, our recent investigation indicates that it can be
applicable for a very general spacetime which is only required to
contain a weakly isolated horizon without
rotation\cite{Wu:2013kqa}. Nevertheless, comparing with the
conventional hydrodynamical expansion method, the method of
imposing Petrov-like boundary condition contains an obvious weak
point, which sticks to the near-horizon limit. Namely, in this
approach we always take the non-relativistical limit with the
near-horizon limit simultaneously. While recently our
understandings on the gravity/fluid duality have been
significantly pushed forward by investigating the hydrodynamical
behavior of gravity at finite cutoff surface based on the
Wilsonian approach or the renormalization group point of
view\cite{Iqbal:2008by,Bredberg:2010ky,Heemskerk:2010hk}. One key
observation in this approach is that any interacting quantum field
theory at finite temperature should be described by hydrodynamics
when viewed at sufficiently long length scales. Since the radial
coordinate $r$ of the AdS bulk corresponds to the energy scale of
the boundary field theory, the near-horizon limit only captures
the low-frequency limit of linear response of the boundary theory
fluid. Therefore, if one intends to move away from the
low-frequency limit, he needs consider fluid membrane at a
hypersurface with constant-radius and finite distance from the
horizon. A flow equation for the radius-dependent response
function has been derived, for instance in \cite{Iqbal:2008by},
which can be viewed as a renormalization group flow to link the
gravity/fluid duality near horizon and that at infinity. In this
paper we intend to establish such duality at finite cutoff surface
with the Petrov-like boundary condition method. Previously in
original Ref.\cite{Bredberg:2011jq} it has been pointed out that
in Rindler spacetime the perturbed metric in bulk obtained by
hydrodynamical expansion may be subject to the Petrov-type
condition at finite cutoff surface(also see the similar check in
\cite{Cai:2013uye}). However, would {\it any} perturbation
constrained by the Petrov-like boundary condition at finite cutoff
surface lead to Navier-Stokes equation? Moreover, can this
observation be extended to more general spacetime background? We
intend to investigate these issues in this paper. We will
demonstrate through explicit models that once the near-horizon
limit is replaced by the long-wavelength limit, the incompressible
Navier-Stokes equation can still be derived by directly imposing
Petrov-like boundary condition on the finite cutoff surface such
that the gravity/fluid duality can be established\footnote{The
need of a long-wavelength limit in finite cutoff case can be
understood from the viewpoint of holography. The radius of the
cutoff corresponds to the energy scale of a dual field theory on
the boundary. In near-horizon case, the energy scale of dual
theory approaches to zero, which means any perturbation of the
dual theory are low energy modes which can be described by
hydrodynamics. However, in finite cutoff case, the energy scale of
dual theory is not low enough such that not all perturbations have
contributions to the hydrodynamical degrees of freedom. In this
sense we need the long-wavelength limit to pick out those low
energy perturbation which corresponds to the degree of freedom of
hydrodynamics.}. Of course in this extension we will only focus on
the cutoff surface which is spatially flat since the
long-wavelength limit is introduced.

To keep this paper in a concise version, we will just present our
main results in the main body, but leave all the detailed
calculation in the appendix.

\section {Petrov-like boundary condition on the finite cutoff
                                 surface for Rindler spacetime}
The framework of imposing Petrov-like boundary condition on the
cutoff surface has been introduced in previous literature, and we
refer to Refs.\cite{Lysov:2011xx,Huang:2011he} for details. Here we
just repeat its basic definition and setting. The Petrov-like
boundary condition on a hypersurface $\Sigma_c$ is defined as
\begin{equation}
    C_{(\ell)i(\ell)j}\equiv\ell^{\mu}{m_i}^{\nu}\ell^{\alpha}{m_j}^{\beta}
    C_{\mu\nu\alpha\beta}=0,
\end{equation}
where $C$ is the Weyl tensor and the Newman-Penrose-like vector
fields satisfy the relations
\begin{equation}
    \ell^2=k^2=0, \ \ (k,\ell)=1, \ \ (k,m_i)=(\ell,m_i)=0,
                \ \ (m^i,m_j)={\delta^i}_j.
\end{equation}
As the simplest example we firstly demonstrate how to derive the
Navier-Stokes equation at finite cutoff surface in Rindler
spacetime. A $p+2$-dimensional metric is
\begin{equation}
    ds^2_{p+2}=-rdt^2+2dtdr+\delta_{ij}dx^idx^j,\ \ \ \ i,j=1,...p.
\end{equation}
Setting $r=r_c$, then we obtain an embedded hypersurface
$\Sigma_c$ and the induced metric $h_{ab}$ on $\Sigma_c$ reads as
\begin{equation}
    ds^2_{p+1}=-r_c dt^2+\delta_{ij}dx^idx^j\equiv -{(dx^0)}^2+
                                           \delta_{ij}dx^idx^j.
\end{equation}
In coordinate system $(t,x^i)$, one can easily check that the
non-vanishing component of the extrinsic curvature $K_{ab}$ is
$K_{tt}=-\sqrt{r_c}/2$. In order to extract the dynamical behavior
of the geometry in the long-wavelength limit as well as the
non-relativistical limit simultaneously, we introduce a parameter
$\lambda$ by rescaling the time coordinate with
$x^0=\frac{1}{\lambda}\tau$ and the space coordinates with
$x^i=\frac{1}{\sqrt{\lambda}}x^I$ such that
\begin{equation}
    ds^2_{p+1}=-\frac{1}{\lambda^2}d\tau^2
       +\frac{1}{\lambda}\delta_{IJ}dx^Idx^J.
\end{equation}
Note that, precisely speaking, here the wavelength is long
compared to the local temperature, which is the natural
characteristic scale in the theory (see, e.g.
\cite{Bhattacharyya:2008ji,Bhattacharyya:2008kq}). Obviously the
non-relativistic limit and long-wavelength limit can be
implemented by taking $\lambda\to 0$.

Next we consider the perturbations of gravity. As we have adopted
before in \cite{Huang:2011he}, \cite{Huang:2011kj} and
\cite{Zhang:2012uy}, we keep the intrinsic metric of the
hypersurface fixed, and then take the Brown-York stress tensor as
the fundamental variable, which is defined as
$t_{ab}\equiv Kh_{ab}-K_{ab}$. In coordinate system $(\tau,x^I)$,
we expand the components of Brown-York tensor in powers of
$\lambda$ as
\begin{eqnarray}
    &&{t^{\tau}}_{\tau}=0+\lambda{{t^{\tau}}_{\tau}}^{(1)}+\dots \nonumber\\
    &&{t^{\tau}}_{I}=0+\lambda{{t^{\tau}}_{I}}^{(1)}+\dots \nonumber\\
    &&{t^{I}}_{J}=\frac{1}{2\sqrt{r_c}}{\delta^I}_J+\lambda{{t^I}_J}^{(1)}
                                                          +\dots \nonumber\\
    &&t=\frac{p}{2\sqrt{r_c}}+\lambda t^{(1)}+\dots, \label{d}
\end{eqnarray}
where $t$ is the trace of Brown-York tensor. In terms of the
Brown-York tensor, the Hamiltonian constraint can be written
as\footnote{The original definitions about the Hamiltonian
constraint and the momentum constraint on the cutoff surface
can be found in \cite{Lysov:2011xx}. Moreover, their
specific forms for the models in our current paper have previously
been presented in \cite{Lysov:2011xx}, \cite{Huang:2011kj} and
\cite{Zhang:2012uy} respectively.}
\begin{equation}
    {({t^{\tau}}_{\tau})}^2
    -\frac{2}{\lambda^2}h^{IJ}{t^{\tau}}_I{t^{\tau}}_J
    +{t^I}_J{t^J}_I-\frac{t^2}{p}=0. \label{f}
\end{equation}
Directly taking the perturbation expansion, we find the
leading order is trivially satisfied by the background
while the sub-leading order with $\lambda^1$ reads as
\begin{equation}
    {{t^{\tau}}_{\tau}}^{(1)}=-2\sqrt{r_c}\delta^{IJ}
             {{t^{\tau}}_I}^{(1)}{{t^{\tau}}_J}^{(1)}.
\end{equation}
Now we turn to the Petrov-like boundary condition. In terms of
the Brown-York tensor in $(\tau,x^I)$, this condition becomes
\begin{equation}
    \lambda{t^{\tau}}_{\tau}{t^I}_J+\frac{2}{\lambda}h^{IK}
    {t^{\tau}}_K{t^{\tau}}_J-2\lambda^2{t^I}_{J,\tau}
    -\lambda{t^I}_K{t^K}_J-2h^{IK}{t^{\tau}}_{(K,J)}
    +\lambda{\delta^I}_J[\frac{t}{p}(\frac{t}{p}
    -{t^{\tau}}_{\tau})+2\lambda\partial_{\tau}\frac{t}{p}]=0.
                                                  \label{g}
\end{equation}
Similarly, one finds the leading order of the expansion is
automatically satisfied by the background, while the sub-leading
order with $\lambda^2$ gives
\begin{equation}
    {{t^I}_J}^{(1)}=2\sqrt{r_c}\delta^{IK}
    {{t^{\tau}}_K}^{(1)}{{t^{\tau}}_J}^{(1)}
    -\sqrt{r_c}\delta^{IK}
    \partial_J{{t^{\tau}}_K}^{(1)}-\sqrt{r_c}
    \delta^{IK}\partial_K{{t^{\tau}}_J}^{(1)}
    +{\delta^I}_J\frac{t^{(1)}}{p}. \label{h}
\end{equation}
Until now we have obtained the sub-leading order of
the Hamiltonian constraint and the Petrov-like boundary condition.
The next step is plugging these results into the momentum
constraint. Its time component and space component will be
identified as the incompressible condition and the Navier-Stokes
equation respectively. Such technical steps have
been used in Refs.\cite{Lysov:2011xx,Huang:2011he,Huang:2011kj}
and \cite{Zhang:2012uy}. Hence, substituting all these results
into the momentum constraint
\begin{equation}
    \partial_a{t^a}_b=0,
\end{equation}
and identifying
\begin{equation}
   {{t^{\tau}}_I}^{(1)}=\frac{1}{2\sqrt{r_c}}v_I, \ \ \ \
   t^{(1)}=\frac{p}{2\sqrt{r_c}}\widetilde{P},
\end{equation}
as the velocity and the pressure fields of the dual fluid,
we obtain the incompressible condition and the Navier-Stokes
equation on a finite cutoff surface as
\begin{eqnarray}
    \partial_Iv^I=0, \\
    \partial_{\tau}v_I+\delta^{JK}v_K\partial_Jv_I
    -\sqrt{r_c}\delta^{JK}\partial_J\partial_Kv_I+
    \partial_I\widetilde{P}=0. \label{ld}
\end{eqnarray}
We find that the kinematic viscosity $\nu_c$ is cutoff dependent
with $\nu_c=\sqrt{r_c}$. First of all, we remark that we have
obtained the same results as those obtained by hydrodynamical
expansion of the metric in
\cite{Bredberg:2011jq}\footnote{Transforming the coordinate system
from $(\tau,x^I)$ to $(t,x^i)$, we easily find the kinematic
viscosity in $(t,x^i)$ is $r_c$ as derived in
\cite{Bredberg:2011jq}. Moreover, we point out that the
Navier-Stokes equation has the same form (or has the same
kinematic viscosity ) in $(\tau,x^I)$ and $(x^0,x^i)$ coordinate
systems, which can be easily proved by dimension analysis and can
be viewed as an alternative representation of the scaling symmetry
of NS equation presented in \cite{Bredberg:2011jq}. Thus we will
always identify our derived equations in $(\tau,x^I)$ system to
those in $(x^0,x^i)$ system as well. }. Secondly, we find that the
previous results obtained for the cutoff surface in near-horizon
limit in \cite{Lysov:2011xx} can be treated as a special case of
our current work. As a matter of fact, transforming the
coordinate system ($\tau,x^I$) into ($\tau,x^i$) which is applied
in \cite{Lysov:2011xx} we find
\begin{equation}
    \partial_\tau v_i+v^j\partial_j
    v_i-\partial^j\partial_j v_i+\partial_iP
    =0, \label{le}
\end{equation}
where $v_i$ is defined as $dx^i/d\tau$ correspondingly.
Therefore, in near-horizon limit one obtains the standard
incompressible Navier-Stokes equation with unit shear viscosity.

\section {Petrov-like boundary condition on the finite cutoff
                          surface for a black brane background}
Next we will treat the Petrov-like boundary condition on the
finite cutoff surface for different backgrounds in a parallel way.
The general framework for Petrov-like boundary condition in
this context is presented in \cite{Huang:2011kj}.
Firstly we consider a black brane background with a metric as
\begin{eqnarray}
    ds^2_{p+2}&=&-f(r)dt^2+2dtdr+r^2{\tilde{\delta}}_{ij}
                           d{\tilde x}^id{\tilde x}^j,
                                 \ \ \ \ \ \ i,j=1,...p,\\
          f(r)&=&r^2(1-\frac{r_h^{p+1}}{r^{p+1}}),
                                  \ \ \ \ \ \ \ \ \ \
        \Lambda=-\frac{p(p+1)}{2}.\nonumber
\end{eqnarray}
Where $r_h$ is the position of the horizon. Setting $r=r_c$, we
have the embedded hypersurface $\Sigma_c$ and its metric reads as
\begin{equation}
    ds^2_{p+1}=-f(r_c)dt^2+{r_c}^2{\tilde{\delta}}_{ij}
    d{\tilde x}^i d{\tilde x}^j\equiv -{(dx^0)}^2
    +\delta_{ij}dx^idx^j.
\end{equation}
It is obvious that this is a intrinsically flat embedding, so that
\begin{equation}
    {^{p+1}\widetilde{R}}_{ij}={^p\widetilde{R}}_{ij}=0.
\end{equation}
Similarly, the non-relativistical and long-wavelength limit is
complemented by rescaling the coordinates as
\begin{equation}
    ds^2_{p+1}=-\frac{1}{\lambda^2}d\tau^2
    +\frac{1}{\lambda}\delta_{IJ}dx^Idx^J.
\end{equation}
Straightforwardly, we take the perturbation expansion for
Brown-York stress tensor and substitute it into the
Petrov-like boundary condition and constraint equations
step by step. We find in this case the sub-leading order
of the Hamiltonian constraint becomes
\begin{equation}
    {{t^{\tau}}_{\tau}}^{(1)}
    =\frac{2\sqrt{f}r_c}{-r_c\partial_{r_c}f+2f}
     \delta^{MN}{{t^{\tau}}_M}^{(1)}{{t^{\tau}}_N}^{(1)}
     +\frac{2f}{-r_c\partial_{r_c}f+2f}t^{(1)}.
\end{equation}
While from the Petrov-like boundary condition we have
\begin{eqnarray}
    {{t^I}_J}^{(1)} &=& \frac{2\sqrt{f}r_c}{r_c\partial_{r_c}f
    +(p-2)f}\delta^{IK}{{t^{\tau}}_K}^{(1)}{{t^{\tau}}_J}^{(1)}
    -\frac{2\sqrt{f}r_c}{r_c\partial_{r_c}f+(p-2)f}\delta^{IK}
    {{t^{\tau}}_{(K,J)}}^{(1)} \nonumber\\
     &&-\frac{f}{r_c\partial_{r_c}f+(p-2)f}
    {\delta^I}_J{{t^{\tau}}_{\tau}}^{(1)}+\frac{r_c\partial_{r_c}f+pf}
    {p[r_c\partial_{r_c}f+(p-2)f]}{\delta^I}_Jt^{(1)}. \label{df}
\end{eqnarray}
 If we identify
\begin{eqnarray}
     &&{{t^{\tau}}_I}^{(1)}
    =\frac{r_c\partial_{r_c}f+(p-2)f}{2\sqrt{f}r_c}v_I, \\
     &&\widetilde{P}
    =\frac{f}{r_c\partial_{r_c}f-2f}
    \delta^{MN}v_Mv_N+\frac{2\sqrt{f}{r_c}^2\partial_{r_c}f}
      {p[r_c\partial_{r_c}f+(p-2)f](r_c\partial_{r_c}f-2f)}t^{(1)},
\end{eqnarray}
then the momentum constraint leads to the incompressible condition
and the Navier-Stokes equation as
\begin{eqnarray}
    \partial_Iv^I=0, \\
    \partial_{\tau}v_I+\delta^{JK}v_K\partial_Jv_I
     -\nu_c\delta^{JK}\partial_J\partial_Kv_I
     +\partial_I\widetilde{P}=0.
\end{eqnarray}
Where $\nu_c=\frac{\sqrt{f}r_c}{r_c\partial_{r_c}f+(p-2)f}$ is the
viscosity of the dual fluid. As argued in previous section, the
form of Navier-Stokes equation will not change when one transforms
the coordinate system to ($x^0,x^i$)
\begin{eqnarray}
    \partial_iv^i=0, \\
    \partial_0v_i+v^j\partial_jv_i
    -\nu_c\partial^j\partial_jv_i+\partial_iP=0,
                               \label{dn}
\end{eqnarray}
with the same viscosity $\nu_c=\frac{\sqrt{f}r_c}
{r_c\partial_{r_c}f+(p-2)f}$.\\
Specially, when
\begin{eqnarray}
    r_c\to r_h, \ \ \ \ \ \ \
    \nu_c=0; \nonumber\\
    r_c\to \infty, \ \ \ \ \ \ \
    \nu_c=\frac{1}{p}. \nonumber
\end{eqnarray}
First of all, after transforming the coordinate system ($x^0,x^i$)
into ($\tau,x^i$), we find that the previous results obtained for
the cutoff surface in near-horizon limit in \cite{Huang:2011kj}
can be treated as a special case of our current work. Secondly,
comparing our results with the previous results presented in
\cite{Cai:2011xv}, the viscosity of both results are equal to zero
when the hypersurface moves to horizon. However, when the
hypersurface moves to infinity, our viscosity approaches to
$\frac{1}{p}$, in contrast to the results in \cite{Cai:2011xv} in
which the viscosity tends to divergence. Such a difference may be
understood from the fact that the Petrov-like boundary condition
we employed in this paper is different from the boundary
conditions in \cite{Cai:2011xv} and \cite{Iqbal:2008by}, which is
the Dirichlet boundary conditions plus a regular condition on the
horizon. Different boundary conditions imply that the dual field
theory that we obtained may be different from that obtained
through the hydrodynamical expansion of the metric. It is
well-known in holography that different boundary conditions lead
to different realizations of holographic duals (for instance, see
\cite{Matsuo:2009pg} and references therein for alternative
realization of Kerr/CFT correspondence). In this sense, we think
we have presented a new way to establish the gravity/fluid duality
at finite cutoff in a black brane background. Finally, we remark
that it should be interesting to investigate the ratio of shear
viscosity to entropy density at finite cutoff in our formalism in
future, which has been found with some universal behavior in
literature\cite{Iqbal:2008by}.

\section {Petrov-like boundary condition on the finite cutoff
                           surface for a background with matter}
The last model is on the gravity/fluid duality in spacetime with
matter fields. The general framework for Petrov-like boundary
condition in this context is presented in \cite{Zhang:2012uy}.
Here we consider a 4-dimensional magnetic black brane, which is a
solution to the Einstein equation coupled to the electromagnetic
field with a metric as
\begin{eqnarray}
    {ds_4}^2&=&-f(r)dt^2+2dtdr+r^2{\tilde\delta}_{ij}d{\tilde x}^id{\tilde x}^j,
                                    \ \ \ \ \ \ i,j=1,2,\\
          f(r)&=&r^2-\frac{2\mu}{r}+\frac{{Q_m}^2}{r^2},
                                     \ \ \ \ \ \ \ \ \
       \Lambda = -3.\nonumber
\end{eqnarray}
Here $\mu$ is the mass parameter and $Q_m$
is the magnetic charge. The electromagnetic field strength is
given by
\begin{equation}
    F=\sqrt{2}Q_md{\tilde x}^1 \wedge d{\tilde x}^2.
\end{equation}
After a straightforward but tedious calculation (the relevant detailed
calculation is presented in the appendix A, B and C), one obtains the
sub-leading order of the Hamiltonian constraint from the expansion
as
\begin{equation}
     {{t^{\tau}}_{\tau}}^{(1)}
    =\frac{2\sqrt{f}r_c}{-r_c\partial_{r_c}f+2f}
      \delta^{MN}{{t^{\tau}}_M}^{(1)}{{t^{\tau}}_N}^{(1)}
      +\frac{2f}{-r_c\partial_{r_c}f+2f}t^{(1)}. \label{za}
\end{equation}
While from the Petrov-like boundary condition, the
sub-leading order of the expansion reads as
\begin{eqnarray}
    {{t^I}_J}^{(1)} &=& \frac{2\sqrt{f}}{\partial_{r_c}f}
    \delta^{IK}{{t^{\tau}}_K}^{(1)}{{t^{\tau}}_J}^{(1)}-
    \frac{2\sqrt{f}}{\partial_{r_c}f}\delta^{IK}
    {{t^{\tau}}_{(K,J)}}^{(1)} \nonumber\\
     &&-\frac{f}{r_c\partial_{r_c}f}
    {\delta^I}_J{{t^{\tau}}_{\tau}}^{(1)}
     +\frac{r_c\partial_{r_c}f+2f}
    {2r_c\partial_{r_c}f}{\delta^I}_Jt^{(1)}. \label{fff}
\end{eqnarray}
In the presence of matter fields, we note the momentum
constraint becomes
\begin{equation}
    \partial_a{t^a}_b=T_{\mu b}n^{\mu},
\end{equation}
where $T_{\mu b}$ is energy-momentum tensor of the matter field.
Similarly, if we identify
\begin{eqnarray}
    &&{{t^{\tau}}_I}^{(1)}=\frac{\partial_{r_c}f}{2\sqrt{f}}v_I, \\
    &&\widetilde{P}=\frac{f}{r_c\partial_{r_c}f-2f}
                   \delta^{MN}v_Mv_N+\frac{\sqrt{f}r_c}
                   {r_c\partial_{r_c}f-2f}t^{(1)},\\
    &&{\widetilde{J}}_J=-\frac{2\sqrt{f}}{\partial_{r_c}f}
                         {F_{nJ}}^{(1)},
\end{eqnarray}
then from the momentum constraint we obtain the incompressible
condition and the  standard incompressible magnetofluid equation
as
\begin{eqnarray}
    \partial_Iv^I=0, \\
    \partial_{\tau}v_I+\delta^{JK}v_K\partial_Jv_I
     -\nu_c\delta^{JK}\partial_J\partial_Kv_I
     +\partial_I\widetilde{P}=f_I.
\end{eqnarray}
Where $\nu_c=\frac{\sqrt{f}}{\partial_{r_c}f}$ is the viscosity
of the dual fluid and $f_I={\widetilde{J}}_J{F_I}^J$ as an external
force term appears on the right hand side of the equation due to
the coupling of the background and the perturbations of the
electromagnetic field.

As argued in previous section, the Navier-Stokes equation has the
same form in coordinate systems ($\tau, x^I$) and ($x^0, x^i$).
Thus, we have the incompressible condition and the standard
incompressible magnetofluid equation in ($x^0, x^i$) as
\begin{eqnarray}
    \partial_iv^i=0, \\
    \partial_0v_i+v^j\partial_jv_i
    -\nu_c\partial^j\partial_jv_i+\partial_iP=f_i,
\end{eqnarray}
with the same viscosity $\nu_c=\frac{\sqrt{f}} {\partial_{r_c}f}$.
Specially, its asymptotic behavior is
\begin{eqnarray}
    r_c\to r_h, \ \ \ \ \ \ \
    \nu_c=0; \nonumber\\
    r_c\to \infty, \ \ \ \ \ \ \
    \nu_c=\frac{1}{2}. \nonumber
\end{eqnarray}

\section{Summary and Discussions}
By explicit construction we have extended the Petrov-like boundary
condition to the finite cutoff surface and derived the
incompressible Navier-Stokes equation in the long-wavelength
limit. In each model, we have computed the value of shear
viscosity and discussed its asymptotical behavior when the
position of cutoff approaches to horizon or infinity. In general
the kinematic viscosity is cutoff dependent and such a dependence
asks for further understanding from the side of holographic
renormalization group flow. In special case when the cutoff
surface approaches to the horizon, our results go back to the
previous ones without employing a long-wavelength limit, implying
a deep analogy between the near-horizon limit and the
long-wavelength limit.

This work, as well as previous works imposing the Petrov-like
boundary condition in the near-horizon limit, only involves the
electromagnetic field as the most simple matter field in the bulk
(see, however, \cite{Wu:2013mda} for the perfect fluid case as a
step further). More general matter fields may lead to further
problems, such as the anisotropy caused by the axion field
\cite{MT}, which is rather interesting. It is also challenging to
extend this framework to a finite cutoff surface which may be
spatially curved. When the spatial part of the hypersurface is
compact, the long-wavelength limit seems not applicable. As
emphasized in \cite{Eling:2009pb}, taking the long-wavelength
limit is essential to reduce the partial differential equation to
ordinary differential equation. However, if the section of cutoff
surface is compact, then the wavelength should have an upper
bound such that the long-wavelength limit can not exist globally.
Nevertheless, for some special non-flat cutoff surface the
long-wavelength limit maybe exist. We leave these issues for
further investigation in future.

\begin{acknowledgments}
Wei Zhang is very grateful to Cheng-Yong Zhang for useful
discussion and help. This work is supported by the Natural Science
Foundation of China under Grant Nos.11175245, 11075206, 11275208,
11178002. Y.Ling also acknowledges the support from Jiangxi young
scientists (JingGang Star) program and 555 talent project of
Jiangxi Province.
\end{acknowledgments}

\section*{Appendix:}
\subsection{The Petrov-like boundary condition in the last model}
From now on, we will present the detailed calculation of the last
model with respect to the gravity/fluid duality in spacetime with
matter fields following the general framework presented in
\cite{Zhang:2012uy}. We have the embedded hypersurface $\Sigma_c$
and its metric reads as
\begin{eqnarray}
    ds^2_{p+1}&=&-f(r_c)dt^2+{r_c}^2{\tilde\delta}_{ij}d{\tilde x}^id{\tilde x}^j
                                                                      \nonumber\\
    &\equiv& -{(dx^0)}^2+\delta_{ij}dx^idx^j \nonumber\\
    &=&-\frac{1}{\lambda^2}d\tau^2
    +\frac{1}{\lambda}\delta_{IJ}dx^Idx^J. \nonumber
\end{eqnarray}
Similarly as we fix the induced metric $h_{ab}$ on the cutoff surface,
we also fix ${F_{ab}|}_{\Sigma_c}$, which could be regarded as the
Dirichlet-like boundary condition. Then we have
\begin{equation}
    {F_{\tau I}|}_{r_c}=0. \nonumber
\end{equation}
${F_n}^b$, ${F_a}^b$ and $F^{ab}$ could be written in terms of
$F_{\mu\nu}$ on $\Sigma_c$ as
\begin{eqnarray}
    &&{{F_n}^{\tau}|}_{r_c}=F_{n\tau}h^{\tau\tau},
                                   \ \ \ \ \ \ \ \ \ \ \
    {{F_n}^I|}_{r_c}=F_{nJ}h^{IJ}, \nonumber\\
    &&{{F_{\tau}}^I|}_{r_c}=F_{\tau J}h^{IJ}=0, \ \ \ \ \
    {{F^{IJ}}|}_{r_c}=F_{KL}h^{KI}h^{LJ}. \nonumber
\end{eqnarray}
Then, the perturbation of electromagnetic field
should take the following form
\begin{eqnarray}
    F_{n\tau}=0+\lambda{F_{n\tau}}^{(1)}, \nonumber\\
    F_{nI}=0+\lambda{F_{nI}}^{(1)}. \nonumber
\end{eqnarray}
Now we will give the detailed calculation from the Petrov-like
boundary condition to equation (\ref{fff}). Firstly we remark that
in the presence of matter fields, the Weyl tensor can be expressed
in terms of the intrinsic curvature and extrinsic curvature as
well as the energy-momentum tensor through Eqs.(3)-(6) in
\cite{Zhang:2012uy}. Moreover, since the extrinsic curvature is
related to the Brown-York stress tensor, we can finally rewrite
the Petrov-like boundary condition in terms of Brown-York stress
tensor as\footnote{Our calculation is applicable for a general
spacetime with matter fields, thus we keep $p$ as general until we
get back to the last model with a magnetic black brane, in which
$p$ is set to $2$.}
\begin{eqnarray}
    &&\lambda{t^{\tau}}_{\tau}{t^I}_J+\frac{2}{\lambda}h^{IK}{t^{\tau}}_K{t^{\tau}}_J
    -2\lambda^2{t^I}_{J,\tau}-\lambda{t^I}_K{t^K}_J-2h^{IK}{t^{\tau}}_{(K,J)}
    +\lambda{\delta^I}_J[\frac{t}{p}(\frac{t}{p}-{t^{\tau}}_{\tau})+
    2\lambda\partial_{\tau}\frac{t}{p}]  \nonumber\\
    &&+\lambda\frac{1}{p}(T_{\delta\beta}n^{\beta}n^{\delta}+2\Lambda+T+{\lambda}^2T_{\tau \tau}
    -2\lambda T_{\delta \tau}n^{\delta}){{\delta}^I}_J-\lambda{T^I}_J=0. \nonumber
\end{eqnarray}
The energy-momentum tensor of electromagnetic field takes the form
\begin{equation}
    T_{\mu\nu}=\frac{1}{4}g_{\mu\nu}F_{\rho\sigma}F^{\rho\sigma}-
                F_{\mu\rho}{F_\nu}^{\rho}. \nonumber
\end{equation}
We have
\begin{eqnarray}
    T_{\delta\beta}n^{\beta}n^{\delta}&=&T_{nn}
      =\frac{1}{4}F_{\rho\sigma}F^{\rho\sigma}
         -F_{n\rho}{F_n}^{\rho}, \nonumber\\
    T&=&\frac{p-2}{4}F_{\rho\sigma}F^{\rho\sigma}, \nonumber\\
    {\lambda}^2T_{\tau\tau}
     &=&-\frac{1}{4}F_{\rho\sigma}F^{\rho\sigma}
        -{\lambda}^2F_{\tau\rho}{F_\tau}^{\rho}, \nonumber\\
    -2\lambda T_{\delta \tau}n^{\delta}
     &=&-2\lambda T_{n\tau}
      =2\lambda F_{n\rho}{F_\tau}^{\rho}, \nonumber\\
    -{T^I}_J&=&-\frac{1}{4}{\delta^I}_JF_{\rho\sigma}F^{\rho\sigma}
                           +F^{I\rho}F_{J\rho}. \nonumber
\end{eqnarray}
The Petrov-like boundary condition further becomes
\begin{eqnarray}
    &&\lambda{t^{\tau}}_{\tau}{t^I}_J+\frac{2}{\lambda}h^{IK}{t^{\tau}}_K{t^{\tau}}_J
    -2\lambda^2{t^I}_{J,\tau}-\lambda{t^I}_K{t^K}_J-2h^{IK}{t^{\tau}}_{(K,J)}
    +\lambda{\delta^I}_J[\frac{t}{p}(\frac{t}{p}-{t^{\tau}}_{\tau})+
    2\lambda\partial_{\tau}\frac{t}{p}]  \nonumber\\
    &&+\lambda\frac{1}{p}(-\frac{1}{2}F_{\rho\sigma}F^{\rho\sigma}-F_{n\rho}{F_n}^{\rho}
    -\lambda^2F_{\tau\rho}{F_\tau}^{\rho}+2\lambda F_{n\rho}{F_\tau}^{\rho}+2\Lambda)
    {\delta^I}_J+\lambda F^{I\rho}F_{J\rho}=0. \nonumber
\end{eqnarray}
Moreover
\begin{eqnarray}
       -\frac{1}{2}F_{\rho\sigma}F^{\rho\sigma}
    &=&-F_{n\tau}F_{n\tau}h^{\tau\tau}-F_{nI}F_{nJ}h^{IJ}
        -\frac{1}{2}F_{IJ}F_{KL}h^{KI}h^{LJ}, \nonumber\\
       -F_{n\rho}{F_n}^{\rho}
    &=&-F_{n\tau}F_{n\tau}h^{\tau\tau}
                           -F_{nI}F_{nJ}h^{IJ},\nonumber\\
       -\lambda^2F_{\tau\rho}{F_\tau}^{\rho}
    &=&-\lambda^2F_{n\tau}F_{n\tau}, \nonumber\\
       2\lambda F_{n\rho}{F_\tau}^{\rho}
    &=&2\lambda(F_{n\tau}{F_\tau}^{\tau}+F_{nI}{F_\tau}^I)
       =0, \nonumber\\
       F^{I\rho}F_{J\rho}
    &=&F_{nJ}F_{nL}h^{IL}+F_{JK}F_{LM}h^{LI}h^{MK}. \nonumber
\end{eqnarray}
So, the Petrov-like boundary condition reads as
\begin{eqnarray}
    &&\lambda{t^{\tau}}_{\tau}{t^I}_J+\frac{2}{\lambda}h^{IK}{t^{\tau}}_K{t^{\tau}}_J
    -2\lambda^2{t^I}_{J,\tau}-\lambda{t^I}_K{t^K}_J-2h^{IK}{t^{\tau}}_{(K,J)}
    +\lambda{\delta^I}_J[\frac{t}{p}(\frac{t}{p}-{t^{\tau}}_{\tau})+
    2\lambda\partial_{\tau}\frac{t}{p}]  \nonumber\\
    &&+\lambda\frac{1}{p}[-2F_{n\tau}F_{n\tau}h^{\tau\tau}-\lambda^2F_{n\tau}F_{n\tau}-2F_{nI}F_{nJ}h^{IJ}
        -\frac{1}{2}F_{IJ}F_{KL}h^{KI}h^{LJ}+2\Lambda]{\delta^I}_J \nonumber\\
    &&+\lambda F_{nJ}F_{nL}h^{IL}+\lambda F_{JK}F_{LM}h^{LI}h^{MK}=0. \nonumber
\end{eqnarray}
After plugging $p=2$ into above equation, we get the
Petrov-like boundary condition
\begin{eqnarray}
    &&\lambda{t^{\tau}}_{\tau}{t^I}_J+\frac{2}{\lambda}h^{IK}{t^{\tau}}_K{t^{\tau}}_J
    -2\lambda^2{t^I}_{J,\tau}-\lambda{t^I}_K{t^K}_J-2h^{IK}{t^{\tau}}_{(K,J)}
    +\lambda{\delta^I}_J[\frac{t}{2}(\frac{t}{2}-{t^{\tau}}_{\tau})+
    \lambda\partial_{\tau}t]  \nonumber\\
    &&+\lambda{\delta^I}_J[-F_{n\tau}F_{n\tau}h^{\tau\tau}-\frac{\lambda^2}{2}F_{n\tau}F_{n\tau}-F_{nI}F_{nJ}h^{IJ}
        -\frac{1}{4}F_{IJ}F_{KL}h^{KI}h^{LJ}+\Lambda] \nonumber\\
    &&+\lambda F_{nJ}F_{nL}h^{IL}+\lambda F_{JK}F_{LM}h^{LI}h^{MK}=0. \nonumber
\end{eqnarray}
Taking the perturbation expansion for Brown-York stress tensor and
electromagnetic field, we find the leading order of the expansion
is automatically satisfied by the background while the sub-leading
order with $\lambda^2$ reads as
\begin{eqnarray}
    {{t^I}_J}^{(1)} &=& \frac{2\sqrt{f}}{\partial_{r_c}f}
    \delta^{IK}{{t^{\tau}}_K}^{(1)}{{t^{\tau}}_J}^{(1)}-
    \frac{2\sqrt{f}}{\partial_{r_c}f}\delta^{IK}
    {{t^{\tau}}_{(K,J)}}^{(1)} \nonumber\\
     &&-\frac{f}{r_c\partial_{r_c}f}
    {\delta^I}_J{{t^{\tau}}_{\tau}}^{(1)}
     +\frac{r_c\partial_{r_c}f+2f}
    {2r_c\partial_{r_c}f}{\delta^I}_Jt^{(1)}. \nonumber
\end{eqnarray}

\subsection{The Hamiltonian constraint in the last model}
Here we give the detailed calculation from the
Hamiltonian constraint to equation (\ref{za}).
The Hamiltonian constraint is
\begin{equation}
    ^{p+1}R+K_{ab}K^{ab}-K^2=2\Lambda+2T_{\mu\nu}n^{\mu}n^{\nu},
     \ \ \ a,b=0,\dots p,\ \ \ \mu,\nu=0,\dots p+1.  \nonumber
\end{equation}
In terms of $t_{ab}=Kh_{ab}-K_{ab}$ in coordinate
system ($\tau, x^I$), we get
\begin{equation}
    {({t^{\tau}}_{\tau})}^2-\frac{2}{\lambda^2}h^{IJ}
    {t^{\tau}}_I{t^{\tau}}_J+{t^I}_J{t^J}_I
    -\frac{t^2}{p}-2\Lambda-2T_{\mu\nu}n^{\mu}n^{\nu}=0.
                                               \nonumber
\end{equation}
Considering the last term on the left-hand side of the above equation
\begin{eqnarray}
    -2T_{\mu\nu}n^{\mu}n^{\nu}=-2T_{nn}
    =F_{n\tau}F_{n\tau}h^{\tau\tau}+F_{nI}F_{nJ}h^{IJ}
        -\frac{1}{2}F_{IJ}F_{KL}h^{KI}h^{LJ}, \nonumber
\end{eqnarray}
then the Hamiltonian constraint becomes
\begin{equation}
    {({t^{\tau}}_{\tau})}^2-\frac{2}{\lambda^2}h^{IJ}
    {t^{\tau}}_I{t^{\tau}}_J+{t^I}_J{t^J}_I-\frac{t^2}{p}
    -2\Lambda+F_{n\tau}F_{n\tau}h^{\tau\tau}+F_{nI}F_{nJ}h^{IJ}
    -\frac{1}{2}F_{IJ}F_{KL}h^{KI}h^{LJ}=0. \nonumber
\end{equation}
Now, considering the perturbation of the electromagnetic field and
meanwhile taking the perturbation expansion for Brown-York stress
tensor, we find the leading order of the expansion is automatically
satisfied by the background while the sub-leading order
with $\lambda^1$ reads as
\begin{equation}
     {{t^{\tau}}_{\tau}}^{(1)}
    =\frac{2\sqrt{f}r_c}{-r_c\partial_{r_c}f+2f}
      \delta^{MN}{{t^{\tau}}_M}^{(1)}{{t^{\tau}}_N}^{(1)}
      +\frac{2f}{-r_c\partial_{r_c}f+2f}t^{(1)}. \nonumber
\end{equation}

\subsection{The momentum constraint in the last model}
Following discussion is about the momentum constraint
\begin{equation}
   \partial_a{t^a}_b=T_{\mu b}n^{\mu}. \nonumber
\end{equation}
The time component of the equation is
\begin{equation}
   \partial_a{t^a}_{\tau}=T_{\mu\tau}n^{\mu}. \nonumber
\end{equation}
Because
\begin{eqnarray}
       \partial_a{t^a}_{\tau}
     &=&\partial_{\tau}{t^\tau}_{\tau}+\partial_I{t^I}_{\tau}
      =\lambda\partial_{\tau}{{t^\tau}_\tau}^{(1)}
        -\frac{1}{\lambda}\partial^I{{t^\tau}_I}^{(1)}+\dots,
                                            \nonumber\\
        T_{\mu\tau}n^{\mu}
     &=&T_{n\tau}=0, \nonumber
\end{eqnarray}
then at leading order it gives rise to
\begin{equation}
    \partial^I{{t^\tau}_I}^{(1)}=0. \nonumber
\end{equation}
The space component of the equation is
\begin{equation}
   \partial_a{t^a}_I=T_{\mu I}n^{\mu}. \nonumber
\end{equation}
Similarly, since
\begin{eqnarray}
       \partial_a{t^a}_I
    &=&\partial_{\tau}{t^\tau}_I+\partial_J{t^J}_I \nonumber\\
    &=&\lambda\partial_{\tau}{{t^\tau}_I}^{(1)}
            +\lambda\partial_J{{t^J}_I}^{(1)}, \nonumber\\
       T_{\mu I}n^{\mu}
    &=&T_{nI} \nonumber\\
    &=&-(0+\lambda{F_{nJ}}^{(1)}){F_I}^J \nonumber\\
    &=&-\lambda{F_{nJ}}^{(1)}{F_I}^J, \nonumber
\end{eqnarray}
then at leading order we have
\begin{equation}
    \partial_\tau{{t^\tau}_I}^{(1)}
    +\partial_J{{t^J}_I}^{(1)}=-{F_{nJ}}^{(1)}{F_I}^J.
                                           \nonumber
\end{equation}

\end{document}